# Spatial Coherence of Thermal Emission by a Sphere

Saman Zare[1] and Sheila Edalatpour[1,2]

[1] Department of Mechanical Engineering, University of Maine, Orono, Maine 04469, USA

[2] Frontier Institute for Research in Sensor Technologies, University of Maine, Orono, Maine 04469, USA

**Abstract**: Analytical expressions for calculating the energy density and spatial correlation function of thermal emission by a homogeneous, isothermal sphere of arbitrary size and material are presented. The spectral distribution and the power law governing the distance-dependent energy density are investigated in the near-field and far-field regimes for silicon carbide (dielectric), silicon (semiconducting) and tungsten (metallic) spheres of various size parameters ranging from $X = 0.002$ to 5. The spatial coherence of thermal field emitted by spheres of different size and material is also studied in both radial and polar directions, and the effect of localized surface phonons (LSPhs) on the correlation length and angle is elucidated. It is shown that the energy density follows a power law of $d^{-2}$ ($d$ is the observation distance) in the far field independent of the size and material of the sphere. The power law in the near field is strongly dependent on the material, size parameter, and the ratio $\frac{d}{a}$ ($a$ is the sphere radius). In the near field, the energy density follows a power law of $d^{-6}$ when $X \ll 1$ and $\frac{d}{a} \gg 1$ (similar to an electric point dipole). With increasing $X$ or decreasing $\frac{d}{a}$, the contribution of multipoles to the energy density increases resulting in an increase in the power of $d$ until the power law converges to that for a semi-infinite medium ($d^{-2.5}$, $d^{-0.5}$, and $d^{-3.5}$ for silicon carbide, silicon and tungsten, respectively, in the intermediate near field, and $d^{-3.5}$, $d^{-3.5}$, and $d^{-2.5}$ for silicon carbide, silicon and tungsten, respectively, in the extreme near field.). It is also found that the spatial correlation length in the radial direction is approximately in the orders of $\lambda$, $0.1\lambda$, and $0.001\lambda$ in the far-field, intermediate near-field, and extreme near-field regimes, respectively, when the multipolar LSPhs are not supported. The correlation angle in the extreme near field is strongly dependent on the sphere size parameter, such that it decreases by three orders of magnitude (from $0.5\pi$ to $0.001\pi$) when $X$ increases from 0.002 to 5. The dependence of the correlation angle on $X$ decreases significantly in the intermediate near-field and far-field regimes, and the correlation angle retains the same order of magnitude ($0.15\pi - 0.7\pi$) for all considered $X$s in these two regimes. While the excitation of dipolar LSPhs does not affect the correlation length and angle of the thermal field, the multipolar LSPhs reduce the spatial coherence in both directions.

# I. INTRODUCTION

Thermal radiation is traditionally considered to be spatially incoherent. However, this is not entirely true as the thermally generated electromagnetic waves can exhibit a high degree of spatial coherence in both near-field ($d \leq \lambda$, where $d$ and $\lambda$ are the observation distance and thermal wavelength, respectively) and far-field ($d > \lambda$) regimes.

While far-field blackbody radiation is spatially correlated over a distance on the order of $\lambda/2$ [1], highly correlated far-field emission can be achieved by engineering materials at the sub-wavelength scale [2-11], for example using gratings [2,3,5,7,11] or multilayers of thin films [4,8,9].

Thermal radiation in the near field can have various coherence behaviors depending on the geometry, material properties, and crystalline structure of the source as well as the observation distance. The coherence of thermal near field is mostly studied within the framework of fluctuational electrodynamics [12], and most of these studies are focused on planar media, i.e., semi-infinite media [1,5,13-15], thin films [16,17], and multilayer hyperbolic media [18]. For a semi-infinite planar source, the spatial coherence of the thermal near field has been studied in the intermediate near-field ($d \leq \lambda$) and extreme near-field ($d \ll \lambda$) regimes. When the source does not support surface polaritons, the coherence length in the intermediate regime has the same order of magnitude as the minimum of the skin depth, $\delta$, and the wavelength [13]. As such, a coherence length much smaller than the thermal wavelength is found for metals which have small skin depths. The excitation of surface polaritons significantly modifies the coherence of the thermal field in the intermediate regime. In this case, the fields are correlated over a distance on the scale of the propagation length of the surface polariton [1,5,13,14]. The coherence length in the extreme near-field regime is much smaller than that in the intermediate regime. In this case, the coherence length is equal to the observation distance from the source, $d$ [1]. The spatial coherence of thermal emission has also been studied for dielectric and metallic thin films [16,17]. In the intermediate regime, the waveguide modes of a dielectric (heavily-doped silicon) thin film result in a long-range coherence ($1.34\lambda$) for the thermal field [16]. In the extreme near-field regime, a very small coherence length ($1.13\times10^{-4}\lambda$) is obtained for the thermal field. The short coherence length of the thermal emission in this regime is attributed to the divergence of the electric field in vicinity of point charges induced at the source surface. The spatial coherence of thermal field emitted by metallic thin films has been studied in the extreme near-field regime [17]. The coherence length in the polariton frequency band increases due to the excitation of surface plasmon polaritons, while the coherence length outside this band is $\lambda/2$. In the polariton frequency band, the coherence length strongly depends on the film thickness and can be larger than $10\lambda$. Spatial coherence of the thermal near field has also been studied for a multilayer hyperbolic ($\varepsilon_\parallel \varepsilon_\perp < 0$, where $\varepsilon_\parallel$ and $\varepsilon_\perp$ are the dielectric functions parallel and perpendicular to the optical axis, respectively) semi-infinite medium [18]. The correlation length of thermal emission by a type I ($\varepsilon_\parallel < 0$ and $\varepsilon_\perp > 0$) hyperbolic semi-infinite medium, which cannot support surface polaritons, is in the order of $0.1\lambda$ and $\lambda$ in the extreme and intermediate near-field regimes, respectively. Unlike type I, a type II ($\varepsilon_\parallel > 0$ and $\varepsilon_\perp < 0$) hyperbolic medium can support surface polaritons resulting in a large correlation length in the order of $10\lambda$ in the intermediate near field. In the extreme near field, however, the correlation length for a type II hyperbolic medium is the same as that for the type I, i.e., in the order of $0.1\lambda$. The non-local effects on the spatial correlation of thermal emission in the extreme near-field regime have been studied for semi-infinite planar media using approximate macroscopic models [15]. The coherence length of the extreme near field is calculated for polar crystals and an electron plasma. For polar crystals, a correlation length equal to the lattice constant is obtained for thermal emission at sub-nanometer distances. For electron plasma, the minimum coherence length is set by the Thomas-Fermi screening length [15].

While spatial coherence of thermal emission by planar media has been extensively studied, spatial coherence for finite, non-planar media is almost unexplored. In the only study concerned with non-planar media [19], the spatial coherence of an incident electric field scattered by a single lossless (non-emitting) sphere with $X = 1$ ($X$ is the size parameter) and a chain of spheres with $X = 0.3$ is studied. The scattering of incident field by the spheres is studied using electromagnetic multiple-scattering theory. The single sphere has a constant (with wavelength) and real dielectric function (thus it does not emit thermal radiation), and it does not support localized surface modes. A coherence length in the order of several wavelengths ($\sim 5\lambda$) is found at the near-field distances from dielectric spheres. It is also shown that spheres with larger dielectric function show a higher degree of coherence. The spatial correlation of thermal field is also studied for a chain of lossless silicon nanospheres with $X = 0.3$ as well as for a chain of metallic nanospheres of the same size with a dielectric function given by the Drude model. In both cases, the coherence length varies significantly with the observation location from the chain.

Spatial coherence of thermal emission by spheres is of significance as they are used for engineering thermal emission in man-made materials such as Mie-resonance-based metamaterials [20-23]. However, the spatial coherence of the thermal field emitted by a single sphere is not fully understood yet. The spatial correlation function for an emitting sphere has not been formulated so far, and the effects of the localized surface phonons (LSPhs), material, and sphere size parameter on spatial coherence in the extreme near-field, intermediate near-field, and far-field regimes have not been investigated yet. Additionally, while the far-field emissive power of spheres with various size parameters are discussed in several studies [24-27], there has not been any analysis of the near-field thermal emission by a sphere.

In this paper, we use the analytical expressions of the electric and magnetic fields thermally emitted by a sphere [24] to calculate the energy density and spatial correlation function. The spectrum and the power laws governing near-field and far-field energy density for spheres of various size parameters and materials are studied. Silicon carbide (SiC), silicon (Si), and tungsten (W) are considered in this study. The spatial coherence of the thermal field emitted by a sphere is also investigated in both radial and polar directions. The effects of size parameter, LSPhs, and material on the spatial coherence of thermal emission are discussed, and the correlation length and angle in the extreme near-field, intermediate near-field, and far-field regimes are quantified.

This paper is organized as follows. Analytical expressions for energy density and spatial correlation function of the thermal field are presented in Section II. These expressions are then used in Section III to study the spectral and total (spectrally-integrated) energy density. The spatial coherence along radial and polar directions is discussed in Section IV, and the concluding remarks are provided in Section V.

## II. MATHEMATICAL DERIVATION OF SPATIAL CORRELATION FUNCTION

The problem under consideration is schematically shown in Fig. 1. A sphere with an arbitrary radius, $a$, and temperature, $T$, is thermally emitting in the free space. The sphere is described by a local, complex, and frequency-dependent dielectric function $\varepsilon = \varepsilon' - i\varepsilon''$ ($e^{i\omega t}$ dependence is used for the time harmonic fields), and it is assumed that the sphere is homogeneous, nonmagnetic, and in local thermodynamic equilibrium. The spatial coherence of thermal electric fields at two arbitrary points $\mathbf{r}_1$ and $\mathbf{r}_2$ located in the free space, which is characterized using the spatial correlation function, is of interest. The spatial correlation function is defined as [16,28,29]:

$$\ddot{\mathbf{W}}(\mathbf{r}_1,\mathbf{r}_2,\omega)\delta(\omega-\omega') = \langle \mathbf{E}(\mathbf{r}_1,\omega) \otimes \mathbf{E}^*(\mathbf{r}_2,\omega') \rangle \tag{1}$$

where $\mathbf{E}(\mathbf{r}_1,\omega)$ and $\mathbf{E}(\mathbf{r}_2,\omega)$ are the thermally-emitted electric fields at points $\mathbf{r}_1$ and $\mathbf{r}_2$, respectively, $\otimes$ denotes the direct product of two vectors, the superscript $*$ represents the complex conjugate, and $\langle \ \rangle$ is the ensemble average. The electric field emitted by the sphere can be obtained using fluctuational electrodynamics [12,24]. The electric field at point $\mathbf{r}$ in the free space can be expanded in vector spherical harmonics as [24,30]:

$$\mathbf{E}(\mathbf{r},\omega) = \sum_{n=1}^{\infty} \sum_{m=-n}^{n} \left( Q_{nm}\mathbf{M}_{nm}(\mathbf{r},\omega) + S_{nm}\mathbf{N}_{nm}(\mathbf{r},\omega) \right) \tag{2}$$

In Eq. (2), $\mathbf{M}_{nm}$ and $\mathbf{N}_{nm}$ are the vector spherical harmonics that are given by the following equations [24]:

$$\mathbf{M}_{nm} = h_n(k_0 r)\left( \frac{im}{\sin\theta} P_n^{|m|} \hat{\theta} - P_n'^{|m|} \hat{\varphi} \right) e^{im\varphi} \tag{3-1}$$

$$\mathbf{N}_{nm} = \left\{ \frac{n(n+1)}{k_0 r} h_n(k_0 r) P_n^{|m|} \hat{r} + \frac{1}{k_0 r} \frac{\partial}{\partial r}\left[r h_n(k_0 r)\right] P_n'^{|m|} \hat{\theta} + \frac{im}{k_0 r \sin\theta} \frac{\partial}{\partial r}\left[r h_n(k_0 r)\right] P_n^{|m|} \hat{\varphi} \right\} e^{im\varphi} \tag{3-2}$$

where $k_0$ is the magnitude of the wavevector in the free space, $h_n$ is the spherical Hankel function of the second order, $i$ is the imaginary unit, $(r,\theta,\varphi)$ are the components of vector $\mathbf{r}$ in spherical coordinates, $\hat{r}$, $\hat{\theta}$, and $\hat{\varphi}$ are the unit vectors in spherical coordinates, $P_n^{|m|}$ is the associated

Legendre polynomial defined as $P_n^{|m|} \equiv P_n^{|m|}(\cos\theta)$, and $P_n'^{|m|} = \dfrac{\partial P_n^{|m|}}{\partial \theta}$. It should be noted that the Gaussian units are used in this paper. The coefficients $Q_{nm}$ and $S_{nm}$ in Eq. (2) are given by [24]:

$$Q_{nm} = -\tilde{A}_{nm} / qa^2 \left[ h_n(k_0 a) j_n'(qa) - h_n'(k_0 a) j_n(qa) \right] \tag{4-1}$$

$$S_{nm} = \tilde{B}_{nm} / k_0 a \left\{ \varepsilon \left[ a h_n(k_0 a) \right]' j_n(qa) - h_n(k_0 a) \left[ a j_n(qa) \right]' \right\} \tag{4-2}$$

where $q = \sqrt{\varepsilon} k_0$, $j_n$ is the spherical Bessel function of the first kind, the prime denotes differentiation with respect to $a$, and the coefficients $\tilde{A}_{nm}$ and $\tilde{B}_{nm}$ are found as [24]:

$$\tilde{A}_{nm} = \int_0^a \frac{k_0(\varepsilon-1)(qr)^2 j_n(qr)}{2\pi \gamma_{nm} \sqrt{\varepsilon}} \int_{-\pi}^{\pi} e^{-im\varphi} d\varphi \int_0^{\pi} \left( \frac{im}{\sin\theta} P_n^{|m|} K_\theta + P_n'^{|m|} K_\varphi \right) \sin\theta \, d\theta \, dr \tag{5-1}$$

$$\begin{aligned}\tilde{B}_{nm}(r) = &-\int_0^a \frac{qk_0 r(\varepsilon-1)}{2\pi \gamma_{nm} \sqrt{\varepsilon}} \frac{\partial (rj_n(qr))}{\partial r} \int_{-\pi}^{\pi} e^{-im\varphi} d\varphi \int_0^{\pi} \left( K_\theta P_n'^{|m|} \sin\theta - imK_\varphi P_n^{|m|} \right) d\theta \, dr \\ &- \int_0^a \frac{(\varepsilon-1)q^2 rj_n(qr)}{2\pi \varepsilon \rho_{nm}} \int_{-\pi}^{\pi} e^{-im\varphi} d\varphi \int_0^{\pi} K_r P_n^{|m|} \sin\theta \, d\theta \, dr \end{aligned} \tag{5-2}$$

where $\gamma_{nm}$ and $\rho_{nm}$ in Eq. (5) are defined as:

$$\gamma_{nm} = \frac{2n(n+1)(n+|m|)!}{(2n+1)(n-|m|)!} \tag{6-1}$$

$$\rho_{nm} = \frac{2(n+|m|)!}{(2n+1)(n-|m|)!} \tag{6-2}$$

In Eq. (5), $\left( K_r, K_\theta, K_\varphi \right)$ are the spherical components of vector $\mathbf{K}$, where $(\varepsilon-1)\mathbf{K}/4\pi$ is the thermally fluctuating electric moment per unit volume [24]. The ensemble average of the spatial correlation function of the spherical components of vector **K** are given by the fluctuation-dissipation theorem as [12,24]:

$$\langle K_\alpha(\mathbf{r},\omega) K^*_\beta(\mathbf{r}',\omega') \rangle = \frac{2\hbar}{r^2 \sin\theta} \coth\left(\frac{\hbar\omega}{2k_B T}\right) \text{Im}\left(\frac{1}{\varepsilon-1}\right) \delta_{\alpha\beta} \delta(\mathbf{r}-\mathbf{r}') \delta(\omega-\omega') \qquad (7)$$

where $\hbar$ and $k_B$ are the reduced Plank and Boltzmann constants, respectively.

Inserting Eq. (2) into Eq. (1) results in:

$$W_{\alpha\beta}(\mathbf{r}_1,\mathbf{r}_2,\omega) = \sum_{n=1}^\infty \sum_{m=-n}^n \sum_{n'=1}^\infty \sum_{m'=-n'}^{n'} \left\langle \left(Q_{nm} M_{nm_\alpha}(\mathbf{r}_1,\omega) + S_{nm} N_{nm_\alpha}(\mathbf{r}_1,\omega)\right) \right. \\ \left. \times \left(Q^*_{n'm'} M^*_{n'm'_\beta}(\mathbf{r}_2,\omega') + S^*_{n'm'} N^*_{n'm'_\beta}(\mathbf{r}_2,\omega')\right) \right\rangle \qquad (8)$$

where $\alpha,\beta = r,\theta,\varphi$. The vector spherical harmonics are deterministic. As such, they can be taken out of the ensemble average, and Eq. (8) can be re-written as follows:

$$W_{\alpha\beta}(\mathbf{r}_1,\mathbf{r}_2,\omega) = \sum_{n=1}^\infty \sum_{m=-n}^n \sum_{n'=1}^\infty \sum_{m'=-n'}^{n'} \left[ \langle Q_{nm} Q^*_{n'm'} \rangle M_{nm_\alpha}(\mathbf{r}_1,\omega) M^*_{n'm'_\beta}(\mathbf{r}_2,\omega') \right. \\ + \langle S_{nm} S^*_{n'm'} \rangle N_{nm_\alpha}(\mathbf{r}_1,\omega) N^*_{n'm'_\beta}(\mathbf{r}_2,\omega') \\ + \langle Q_{nm} S^*_{n'm'} \rangle M_{nm_\alpha}(\mathbf{r}_1,\omega) N^*_{n'm'_\beta}(\mathbf{r}_2,\omega') \\ \left. + \langle S_{nm} Q^*_{n'm'} \rangle N_{nm_\alpha}(\mathbf{r}_1,\omega) M^*_{n'm'_\beta}(\mathbf{r}_2,\omega') \right] \qquad (9)$$

The ensemble average of the products of coefficients $S$ and $Q$ in Eq. (9) are obtained using the following equations:

$$\langle Q_{nm} Q^*_{n'm'} \rangle = \frac{k_0^2 \hbar}{\pi \gamma_{nm} |D_n|^2} \coth\left(\frac{\hbar\omega}{2k_B T}\right) \text{Im}\left(j^*_n(qa) j'_n(qa)\right) \delta_{nn'} \delta_{mm'} \qquad (10\text{-}1)$$

$$\langle S_{nm} S^*_{n'm'} \rangle = \frac{k_0^2 \hbar}{\pi \gamma_{nm} |E_n|^2} \coth\left(\frac{\hbar\omega}{2k_B T}\right) \text{Im}\left\{\varepsilon^* \left(\frac{|j_n(qa)|^2}{a} + j^*_n(qa) j'_n(qa)\right)\right\} \delta_{nn'} \delta_{mm'} \qquad (10\text{-}2)$$

$$\langle Q_{nm} S^*_{n'm'} \rangle = \langle S_{nm} Q^*_{n'm'} \rangle = 0 \tag{10-3}$$

where

$$D_n = a\left[ h_n(k_0 a) j'_n(qa) - h'_n(k_0 a) j_n(qa) \right] \tag{11-1}$$

$$E_n = \varepsilon \left[ a h_n(k_0 a) \right]' j_n(qa) - h_n(k_0 a) \left[ a j_n(qa) \right]' \tag{11-2}$$

Using Eq. (10-3) and the properties of the Kronecker delta function in Eqs. (10-1) and (10-2), the correlation function in Eq. (9) is reduced to:

$$W_{\alpha\beta}(\mathbf{r}_1, \mathbf{r}_2, \omega) = \sum_{n=1}^{\infty} \sum_{m=-n}^{n} \left[ \langle |Q_{nm}|^2 \rangle M_{nm_\alpha}(\mathbf{r}_1, \omega) M^*_{nm_\beta}(\mathbf{r}_2, \omega') \right. \\ \left. + \langle |S_{nm}|^2 \rangle N_{nm_\alpha}(\mathbf{r}_1, \omega) N^*_{nm_\beta}(\mathbf{r}_2, \omega') \right] \tag{12}$$

Equation (12) combined with Eqs. (3) and (10) provide the spatial correlation function of the electric fields at points $\mathbf{r}_1$ and $\mathbf{r}_2$. These equations can also be used for finding the energy density emitted by the sphere in the free space. The electric energy density, $u^E$, at point $\mathbf{r}$ in the free space is given by [31]:

$$u^E(\mathbf{r}, \omega) = \frac{1}{8\pi} \langle |\mathbf{E}(\mathbf{r}, \omega)|^2 \rangle \tag{13}$$

where based on Eq. (1), $\langle |\mathbf{E}(\mathbf{r}, \omega)|^2 \rangle$ can be obtained by taking the trace of $\ddot{\mathbf{W}}(\mathbf{r}_1, \mathbf{r}_2, \omega)$ for $\mathbf{r}_1 = \mathbf{r}_2 = \mathbf{r}$, i.e.,

$$u^E(\mathbf{r}, \omega) = \frac{1}{8\pi} Tr\left[ \ddot{\mathbf{W}}(\mathbf{r}, \mathbf{r}, \omega) \right] \tag{14}$$

Substituting for $\ddot{\mathbf{W}}$ from Eq. (12), the electric energy density can then be written as:

$$u^E(\mathbf{r},\omega) = \frac{1}{8\pi} \sum_{\alpha=r,\theta,\varphi} \sum_{n=1}^{\infty} \sum_{m=-n}^{n} \left[ \langle |Q_{nm}|^2 \rangle |M_{nm_\alpha}(\mathbf{r},\omega)|^2 + \langle |S_{nm}|^2 \rangle |N_{nm_\alpha}(\mathbf{r},\omega)|^2 \right] \quad (15)$$

The magnetic energy density, $u^H$, can be obtained in a similar way as the electric energy density. The magnetic energy density at point **r** in the free space is found using the magnetic field at this point as [31]:

$$u^H(\mathbf{r},\omega) = \frac{1}{8\pi} \langle |\mathbf{H}(\mathbf{r},\omega)|^2 \rangle \quad (16)$$

The magnetic field at point **r** can be expanded in vector spherical harmonics as [24]:

$$\mathbf{H}(\mathbf{r},\omega) = i \sum_{n=1}^{\infty} \sum_{m=-n}^{n} \left( S_{nm} \mathbf{M}_{nm}(\mathbf{r},\omega) + Q_{nm} \mathbf{N}_{nm}(\mathbf{r},\omega) \right) \quad (17)$$

Using Eq. (17) and following the same steps as those taken for deriving the electric energy density, the magnetic energy density is obtained as:

$$u^H(\mathbf{r},\omega) = \frac{1}{8\pi} \sum_{\alpha=r,\theta,\varphi} \sum_{n=1}^{\infty} \sum_{m=-n}^{n} \left[ \langle |S_{nm}|^2 \rangle |M_{nm_\alpha}(\mathbf{r},\omega)|^2 + \langle |Q_{nm}|^2 \rangle |N_{nm_\alpha}(\mathbf{r},\omega)|^2 \right] \quad (18)$$

Finally, the spectral energy density is found by adding Eq. (18) and Eq. (15) as:

$$u(\mathbf{r},\omega) = \frac{1}{8\pi} \sum_{\alpha=r,\theta,\varphi} \sum_{n=1}^{\infty} \sum_{m=-n}^{n} \left( \langle |S_{nm}|^2 \rangle + \langle |Q_{nm}|^2 \rangle \right) \left( |M_{nm_\alpha}(\mathbf{r},\omega)|^2 + |N_{nm_\alpha}(\mathbf{r},\omega)|^2 \right) \quad (19)$$

## III. ENERGY DENSITY

The total (spectrally-integrated) and the spectral energy density emitted by single spheres made of SiC (dielectric), Si (semiconductor), and W (metal) is calculated at an observation distance, $d = r - a$, using the formalism described in Section II. The total energy density is shown in Fig. 2 versus the observation distance normalized by the dominant wavelength of thermal radiation at 300 K ($\lambda_{max} = 9.66$ μm), $\frac{d}{\lambda_{max}}$, for four size parameters ($X = \frac{2\pi a}{\lambda_{max}}$) of 0.002, 0.02, 0.2, and 5. The energy density emitted by a semi-infinite medium [14] as well as an electric point dipole [14] is also shown in Fig. 2 for comparison. The energy density for all materials increases with increasing the size parameter and decreasing the observation distance.

In the far-field region ($\frac{d}{\lambda_{max}} > 10$), the spheres behave as electric point dipoles when their size is much smaller than the wavelength inside the material (i.e., $\lambda/n$ where $n$ is the refractive index of the sphere). As shown in Fig. 2, the far-field energy density for SiC and Si spheres agrees well with that predicted using the dipole approximation when $X < 1$. For W spheres, which have a large refractive index, the agreement between the energy density of the sphere and the dipole approximation is observed in the far field only for $X = 0.002$. The energy density for all spheres follows a power law of $d^{-2}$ in the far field ($\frac{d}{\lambda_{max}} > 10$) regardless of the size and material of the sphere, which is similar to the far-field behavior of an electric point dipole. The electric field generated by a dipole is proportional to $d^{-1}$ in the far field [32]. Consequently, based on Eq. 13 (magnetic energy density is negligible), the energy density of an electric dipole is proportional to $d^{-2}$ in the far field.

In the intermediate near-field distances ($0.01 \leq \frac{d}{\lambda_{max}} \leq 0.1$), the energy density of a sphere with $X$ = 0.002 varies with distance as $d^{-6}$ similar to the behavior of an electric point dipole. For an electric dipole, the electric field follows a power law of $d^{-3}$ in the intermediate near field [32,33], and thus the energy density has a $d^{-6}$ distance dependence. This similarity to a dipole in distance dependence is because $X \ll 1$ and $a \ll d$ in the intermediate near-field distances from a sphere with $X = 0.002$. As the size of the sphere increases, the energy density starts deviating from that of an electric dipole. In this region, the energy density is governed by a $d^{-5}$ power law for SiC, Si, and W spheres with $X = 0.02$. With further increase in the size, the power of $d$ becomes material-dependent and further increases until it converges to that for a semi-infinite medium ($X \to \infty$) when $X = 5$. For $X = 5$, the power laws are $d^{-2.5}$, $d^{-0.5}$, and $d^{-3.5}$ for SiC, Si, and W spheres, respectively.

In the extreme near-field region ($\frac{d}{\lambda_{max}} < 0.01$), the behavior of the energy density for a sphere with $X = 0.002$, where still $a < d$, is similar to that for a point dipole and the energy density is governed by a power law of $d^{-5}$ for all three materials. With increasing the size parameter, however, the size of the spheres becomes comparable to or greater than $d$. As such, the sphere does not behave as a point dipole anymore. Similar to the intermediate near-field regime, the power of $d$ becomes material-dependent and increases with increasing the size of the sphere. As shown in Fig. 2, the energy density and its distance dependence eventually approach those for a semi-infinite medium in the extreme near field ($d^{-2}$ for metals and $d^{-3}$ for dielectrics and semiconductors [1]).

The spectral energy density emitted by a SiC sphere in the extreme near-field, intermediate near-field, and far-field regimes is shown in Fig. 3(a) to 3(c), respectively, for various size parameters ranging from 0.002 to 5. Spheres with sizes smaller than or comparable to the wavelength can support LSPh modes when $\text{Re}[\varepsilon] \approx -(l+1)/l$, where $l = 1, 2, 3,...$ is the order of the mode [34]. The excitation of LSPhs increases thermal emission resonantly resulting in sharp peaks in the energy density. The peaks in the energy density spectra of Fig. 3 are labeled with the order of the associated LSPh modes. As it is seen from Figs. 3(a) and (3b), at a given near-field distance, the order of the excited LSPh modes increases from $l = 1$ (dipole mode) to $l \to \infty$ (surface phonon-polariton mode of a semi-infinite medium) as the size parameter increases from $X = 0.002$ to $X = 5$. Also, the higher order modes become more dominant in the energy density spectra as $X$ increases. However, in the far-field regime (Fig. 3(c)), the dipole mode dominates the energy density regardless of the size when $X \leq 1$ (i.e., when the sphere is smaller or comparable to the wavelength). When $X > 1$, the LSPh peaks are suppressed and broadened due to the retardation effects [35,36]. In addition to the LSPh modes, large electrical losses (i.e., large values of $\text{Im}[\varepsilon]$) can result in peaks in the energy density when the sphere size is comparable to or larger than the wavelength. The spectral energy density for the Si and W spheres does not show any peaks and mostly follows the spectrum of a blackbody.

## IV. SPATIAL COHERENCE

The formalism developed in Section II is used to study the spatial coherence of thermal radiation by a single sphere for various materials, observation distances, and size parameters. The correlation function matrix, $\ddot{\mathbf{W}}$, is calculated for two points ($\mathbf{r}_1 = (r_1, \theta_1, \varphi_1)$ and $\mathbf{r}_2 = (r_2, \theta_2, \varphi_2)$) along the radial direction, i.e., when $\theta_1 = \theta_2 = \theta$ and $\varphi_1 = \varphi_2 = \varphi$ (see the inset of Fig. 4(a)) as well as along the polar direction, i.e., when $r_1 = r_2 = r$ and $\varphi_1 = \varphi_2 = \varphi$ (see the inset of Fig. 8(a)). The spatial correlation of thermal emission is calculated at three observation distances ($d = r_1 - a$) of $\frac{d}{\lambda_{max}} = 0.001$ (located in the extreme near field), $\frac{d}{\lambda_{max}} = 0.1$ (located in the intermediate near field), and $\frac{d}{\lambda_{max}} = 10$ (located in the far field). While point $\mathbf{r}_1$ is fixed at one of these observation locations, the distance of point $\mathbf{r}_2$ from $\mathbf{r}_1$ is increased until the correlation function drops to negligible values. The components of the correlation function are normalized by their values at $\mathbf{r}_1 = \mathbf{r}_2$. The correlation function is first calculated for a wavelength of $\lambda = 11$ µm at which localized surface phonons (LSPhs) are not excited. Then, the effect of the LSPhs on the coherence of thermal emission is discussed.

### A. *Spatial correlation along radial direction*

Spatial correlation along the radial direction is studied in this sub-section. Because of the spherical symmetry in this case, any arbitrary value can be assigned to $\theta$ and $\varphi$. In our calculations, $\theta = \frac{\pi}{2}$ and $\varphi = 0$ are selected (see the inset of Fig. 4(a)). It should also be mentioned that $W_{\theta\theta}$ and $W_{\varphi\varphi}$ are equal due to the spherical symmetry.

Figure 4 shows the diagonal components of the spatial correlation function in the extreme near-field regime. Figure 4(a) compares radial ($W_{rr}$) and polar ($W_{\theta\theta}$) components of the correlation function for a SiC sphere with $X = 0.002$. The radial and polar components are the same, and both exhibit an extremely short spatial correlation length in the extreme near field such that they decay rapidly within a distance of about $0.002\lambda$. The equality of $W_{\theta\theta}$ and $W_{rr}$ are also observed for $X = 0.2$ and 5 (the results are not shown). The polar component ($W_{\theta\theta}$) for a SiC sphere is compared for various size parameters in Fig. 4(b). In this region, the correlation function of the SiC sphere with $X = 0.002$ is similar to that of a dipole. The SiC spheres with $X = 0.2$ and 5 show larger correlation lengths compared to the sphere with $X = 0.002$. The correlation function for these two size parameters converge to the one for a semi-infinite medium in the extreme near field. Still, the spatial correlation length for spheres with $X = 0.2$ and 5 is limited to a small fraction of wavelength ($\sim 0.006\lambda$) in the extreme near field. The polar component of correlation function for a sphere with $X = 0.2$ is compared for different materials in Fig. 4(c). As it is seen from this figure, the spatial coherence is material independent. To summarize, the correlation length at the extreme near-field distances from the sphere is very short (a few one-thousandth of wavelength), slightly increases with the size of the sphere, and does not depend on the material.

Figure 5 shows the normalized correlation function in the intermediate near field. In this regime, the polar component of the correlation function shows a longer spatial correlation length than the radial component for a SiC sphere with $X = 0.002$, as shown in Fig. 5(a). The same behavior is observed for $X = 0.2$ and 5 (the results are not shown). Also, the length of coherence in this case is in the order of $\sim 0.1\lambda$ which is much longer than that in the extreme near field. Similar to the extreme near-field region, the spatial correlation for the sphere with $X = 0.002$ in the intermediate near field is similar to the one for a dipole. However, the enhancement of spatial correlation with increasing the size parameter is more significant in the intermediate near field than in the extreme near field (Fig. 5(b)), such that the spatial coherence for SiC spheres with $X = 5$ extends to distances in the order of a wavelength ($\sim \lambda$). Also, Fig. 5(c) shows that Si and SiC spheres with $X = 0.2$ exhibit the same spatial coherence in the intermediate near field, while the correlation function decays more rapidly for a W sphere. To summarize, in the intermediate near field, the correlation length is significantly longer than that in the extreme near field (between two and three orders of magnitude depending on the size parameter). Additionally, the spatial correlation is material-dependent, and it increases with increasing the size of the sphere.

The normalized correlation function in the far-field regime is plotted in Fig. 6. As illustrated in Fig. 6(a), the polar component of the correlation function is larger than the radial component. Also, the coherence length of thermal radiation is significantly larger than that in the intermediate and extreme near fields, such that it extends to distances in the order of $10\lambda$. In addition, the spatial correlation agrees with the dipole approximation in the far field and does not vary with size parameter (Fig. 6(b)) and material (Fig. 6(c)).

To quantify the distance up to which the thermal field is correlated, the correlation length is calculated. The correlation length is defined as [16]:

$$L_{cor} = \frac{2\int_0^\infty \left|Tr\left(\ddot{\mathbf{W}}(\mathbf{r}_1,\mathbf{r}_2,\omega)\right)\right|^2 dx}{\left|Tr\left(\ddot{\mathbf{W}}(\mathbf{r}_1,\mathbf{r}_1,\omega)\right)\right|^2} \qquad (19)$$

where $x = |r_2 - r_1|$. The correlation length normalized by the wavelength is calculated for a SiC sphere with different size parameters in the three regimes and is plotted versus wavelength in Fig. 7. It is observed that the correlation length significantly increases with increasing the observation distance from the sphere for all size parameters. In addition, the correlation length monotonically decreases with the wavelength in the three regimes and for all size parameters except for a few wavelengths at which the correlation length has local minima and maxima. The wavelengths associated with LSPhs and large electrical losses, which result in enhanced thermal emission as shown in Figs. 3(a)-(c), are marked in Figs. 7(a)-(d). It can be seen that the excitation of the dipolar LSPhs ($l$ = 1) of the sphere does not affect the correlation length while the presence of the multipolar LSPhs ($l > 1$) reduces the spatial correlation in the extreme and intermediate near fields. With increasing the size parameter, the correlation length in the extreme and intermediate near field converges to that for a semi-infinite medium. Furthermore, the modes resulting from large losses ($\text{Im}[\varepsilon] \to \infty$) in the sphere increase the correlation length of the thermal field.

*B. Spatial correlation along polar direction*

Spatial correlation along the polar direction is studied in this sub-section. In this case, points $\mathbf{r}_1$ and $\mathbf{r}_2$ are located at the same radial distance and azimuthal angle (i.e., $r_1 = r_2 = r$ and $\varphi_1 = \varphi_2 = \varphi$) but have different polar angles (see the inset of Fig. 8(a)). In the calculations, $\theta_1 = \frac{\pi}{2}$ and $\varphi = 0$, and three different values of $r$ located in the extreme near-field, intermediate near-field, and far-field regimes are considered.

The normalized correlation function at an extreme near-field distance of $\frac{d}{\lambda_{max}} = 0.001$ is plotted versus $\frac{|\theta_2 - \theta_1|}{\pi}$ in Fig. 8. Figure 8(a) shows the diagonal components of the spatial correlation function for a SiC sphere with $X = 0.002$. As it is seen from this figure, the azimuthal component of the correlation function ($W_{\varphi\varphi}$) exhibits strong spatial coherence for all values of $\frac{|\theta_2 - \theta_1|}{\pi}$, while the radial ($W_{rr}$) and polar ($W_{\theta\theta}$) components decrease slowly with $\frac{|\theta_2 - \theta_1|}{\pi}$ and become negligible at $\frac{|\theta_2 - \theta_1|}{\pi} = 0.5$. The effect of size parameter on $W_{\theta\theta}$ at the extreme near field is shown in Fig. 8(b). It is seen that $W_{\theta\theta}$ decreases significantly with increasing the size of the sphere, such that it vanishes for the sphere with $X = 5$ within a separation angle ($|\theta_2 - \theta_1|$) of only about $0.01\pi$. The same size dependence is observed for $W_{rr}$ and $W_{\varphi\varphi}$ (the results are not shown). Figure 8(c) compares $W_{\theta\theta}$ for SiC, Si, and W spheres with $X = 0.2$. This figure shows that the correlation function in the extreme near field does not depend on the material of the sphere. In summary, the spatial correlation in the extreme near field decreases rapidly with increasing the size parameter and is material independent.

The spatial coherence shows similar behavior in the intermediate near-field ($\frac{d}{\lambda_{max}} = 0.1$) and far-field ($\frac{d}{\lambda_{max}} = 10$) regimes as shown in Figs. 9 and 10, respectively. As it is seen from Figs. 9(a) and 10(a), $W_{\varphi\varphi}$ is equal to unity at all separation angles, while $W_{rr}$ and $W_{\theta\theta}$ (which are equal) slowly decrease to zero when $\frac{|\theta_2 - \theta_1|}{\pi} = 0.5$. The size dependence of spatial coherence in these two regimes is studied using Fig. 9(b) and 10(b), respectively. Unlike in the extreme near-field regime, the spatial coherence of the SiC sphere does not drastically decrease by increasing the size parameter in the intermediate near-field and far-field regimes. Figures 9(c) and 10(c) compare the spatial correlation for a sphere with $X = 0.2$ made of SiC, Si, and W. The correlation function for the SiC sphere is very close to that for the Si sphere, and it decreases from 1 at $\frac{|\theta_2 - \theta_1|}{\pi} = 0$ to 0 at $\frac{|\theta_2 - \theta_1|}{\pi} = 0.5$. The W sphere shows a remarkably stronger spatial coherence along the polar angle, such that the spatial correlation for W sphere is almost equal to 1 for all values of $\frac{|\theta_2 - \theta_1|}{\pi}$. To summarize, the spatial correlation at the intermediate near-field and far-field distances is greater than that in the extreme near field, and it is material-dependent.

Analogous to the length of correlation, we define a correlation angle for the spatial coherence as:

$$\Theta_{cor} = \frac{2\int_0^{\pi/2} \left|Tr\left(\ddot{\mathbf{W}}(\mathbf{r}_1,\mathbf{r}_2,\omega)\right)\right|^2 dx}{\left|Tr\left(\ddot{\mathbf{W}}(\mathbf{r}_1,\mathbf{r}_1,\omega)\right)\right|^2} \tag{20}$$

where $x$ is defined as $x = |\theta_2 - \theta_1|$. The correlation angle is plotted versus wavelength for SiC spheres with $X = 0.002$, $0.02$, $0.2$, and $5$ in Figs. 11(a) to 11(d), respectively. The wavelengths at which energy density has peaks due to the excitation of LSPhs or the presence of large losses is marked in Figs. 11(a)-(d). It can be seen that the effect of LSPhs and large losses on the correlation angle is the same as that on the correlation length along the radial direction. While the dipolar LSPhs ($l = 1$) does not significantly affect the correlation angle, the multipolar LSPhs ($l > 1$) reduce the spatial correlation in the extreme and intermediate near-field regimes. Additionally, the modes due to the large losses in the sphere ($\text{Im}[\varepsilon] \to \infty$) increase the correlation angle in the intermediate near field.

## V. CONCLUSIONS

The spectrum and spatial coherence of thermal emission by a homogeneous and isothermal sphere were analyzed in the extreme near-field, intermediate near-field, and far-field regimes. The energy density and correlation function of a sphere with an arbitrary size parameter and material were formulated using fluctuational electrodynamics. The energy density and spatial coherence along the radial and polar directions were studied for SiC, Si, and W spheres with $X$ ranging from 0.002 to 5. It was shown that the energy density follows a power law of $d^{-2}$ in the far field regardless of the size and material of the sphere. In the near-field regime, the energy density is governed by a power law of $d^{-6}$ when $X \ll 1$ and $\dfrac{d}{a} \gg 1$ (dipole approximation is valid). The power of $d$ in the near field increases with increasing $X$ and decreasing $\dfrac{d}{a}$ until it eventually converges to that for a semi-infinite medium (between $d^{-0.5}$ and $d^{-3.5}$ depending on material and the near-field regime). It was also shown that the correlation length in the radial direction increases by several orders of magnitude from $\sim 0.001\lambda$ in the extreme near field to $\sim 0.1\lambda$ and $\sim \lambda$ in the intermediate near field and far field, respectively. The correlation angle in the extreme near field is strongly size-dependent varying from $0.5\pi$ to $0.001\pi$ when $X$ increases from 0.002 to 5. In the intermediate near-field and far-field regimes, the correlation angle has an order of magnitude of $\sim 0.1\pi$. It was demonstrated that the multipolar LSPhs reduce the spatial coherence along both radial and polar directions in the extreme and intermediate near-field regimes, whereas the dipolar LSPhs do not have any impacts on the spatial coherence. The spectral and spatial properties of thermal emission by a sphere are of great importance since spheres are the building blocks for many man-made materials.

**Acknowledgments**

The authors acknowledge support from the National Science Foundation under Grant No. CBET-1804360.

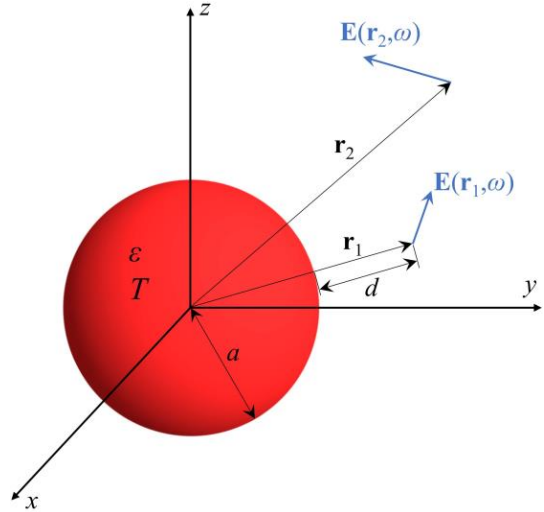

Figure 1 – A homogeneous sphere of radius *a*, temperature *T*, and dielectric function ε is emitting in the free space. The spatial correlation of the thermal electric fields at points $\mathbf{r}_1$, $\mathbf{E}(\mathbf{r}_1,\omega)$, and $\mathbf{r}_2$, $\mathbf{E}(\mathbf{r}_2,\omega)$, is of interest. Parameter *d* shows the distance of observation point $\mathbf{r}_1$ from the sphere surface.

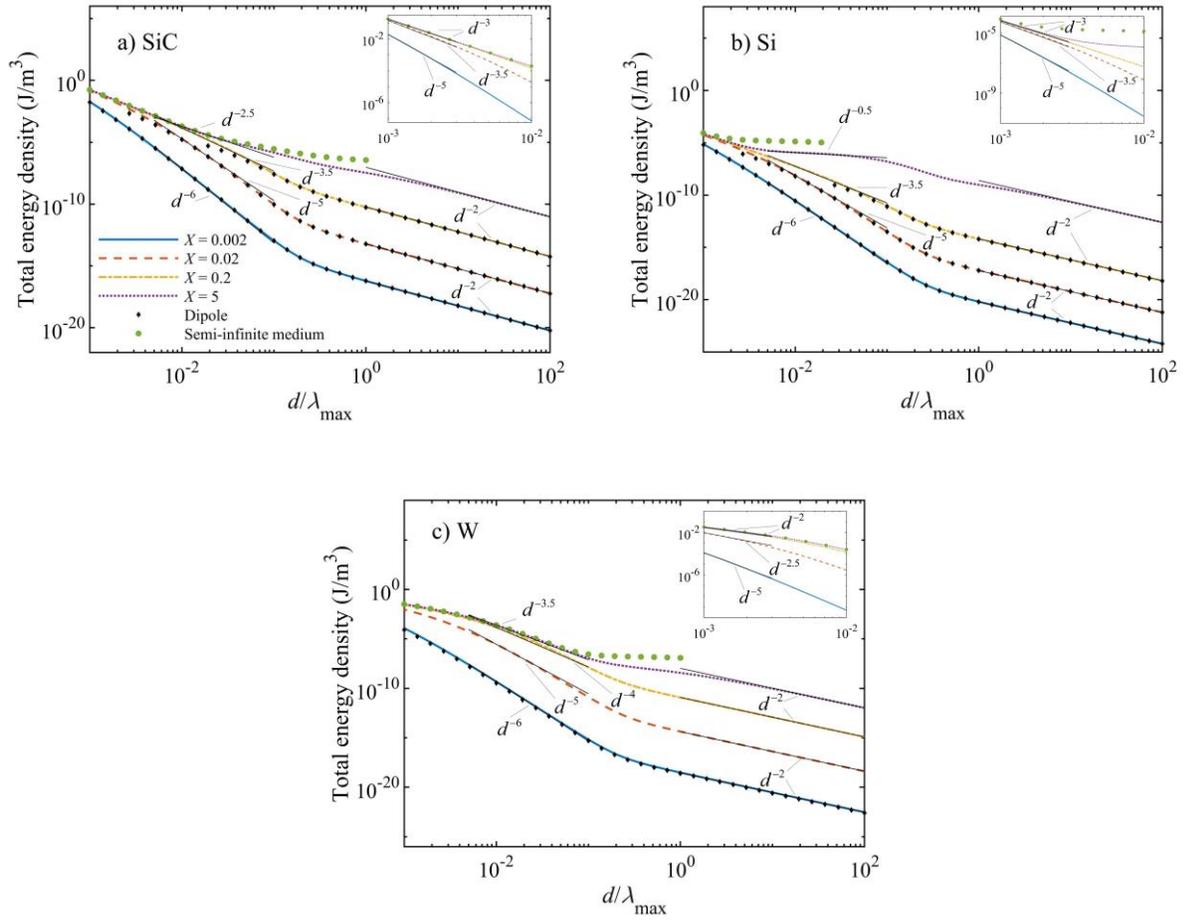

Figure 2 – Total energy density emitted by a (a) SiC, (b) Si, and (c) W sphere versus observation distance $d$ normalized by $\lambda_{max}$ (= 9.66 μm). The insets show the total energy density in the extreme near-field distances ($\frac{d}{\lambda_{max}} < 0.01$).

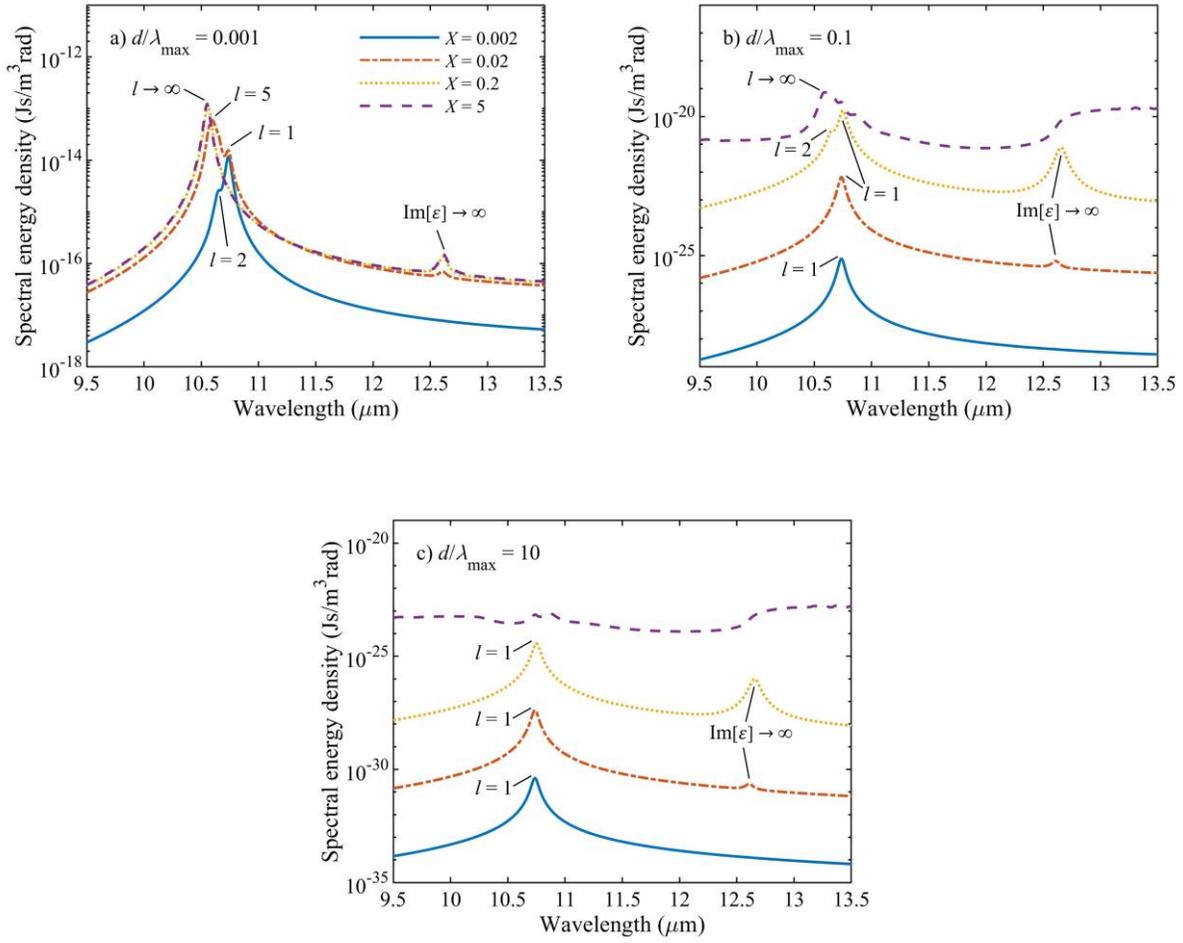

Figure 3 – Spectral energy density emitted by a SiC sphere with various size parameters in the (a) extreme near-field ($\frac{d}{\lambda_{max}} = 0.001$), (b) intermediate near-field ($\frac{d}{\lambda_{max}} = 0.1$), and (c) far-field ($\frac{d}{\lambda_{max}} = 10$) regimes.

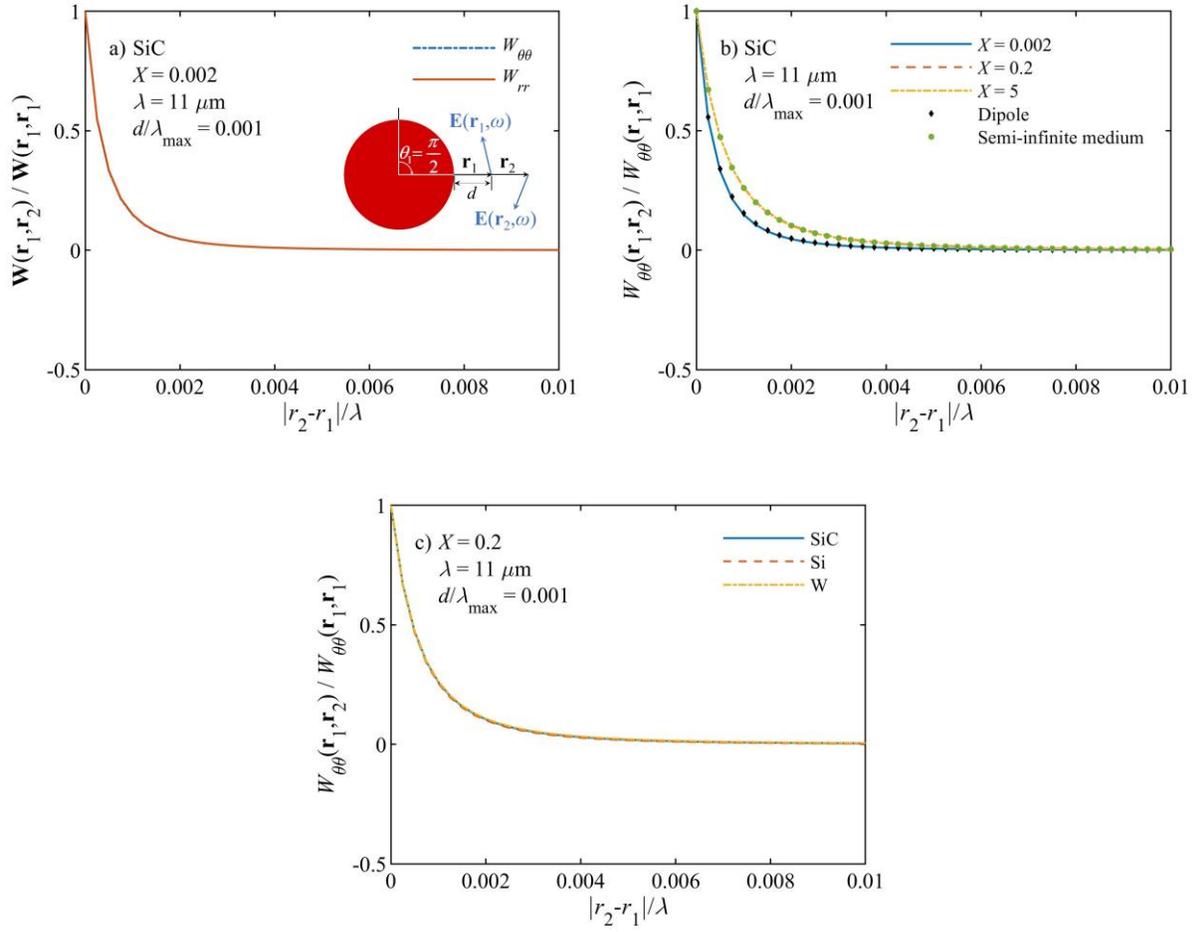

Figure 4 – Normalized correlation function along radial axis in the extreme near-field regime ($\frac{d}{\lambda_{max}} = 0.001$) versus $\frac{|r_2 - r_1|}{\lambda}$ at $\lambda = 11$ µm. (a) Normalized $W_{\theta\theta}$ and $W_{rr}$ for a SiC sphere with $X = 0.002$. (b) Normalized $W_{\theta\theta}$ for SiC spheres with various size parameters. (c) Normalized $W_{\theta\theta}$ for SiC, Si, and W spheres with $X = 0.2$.

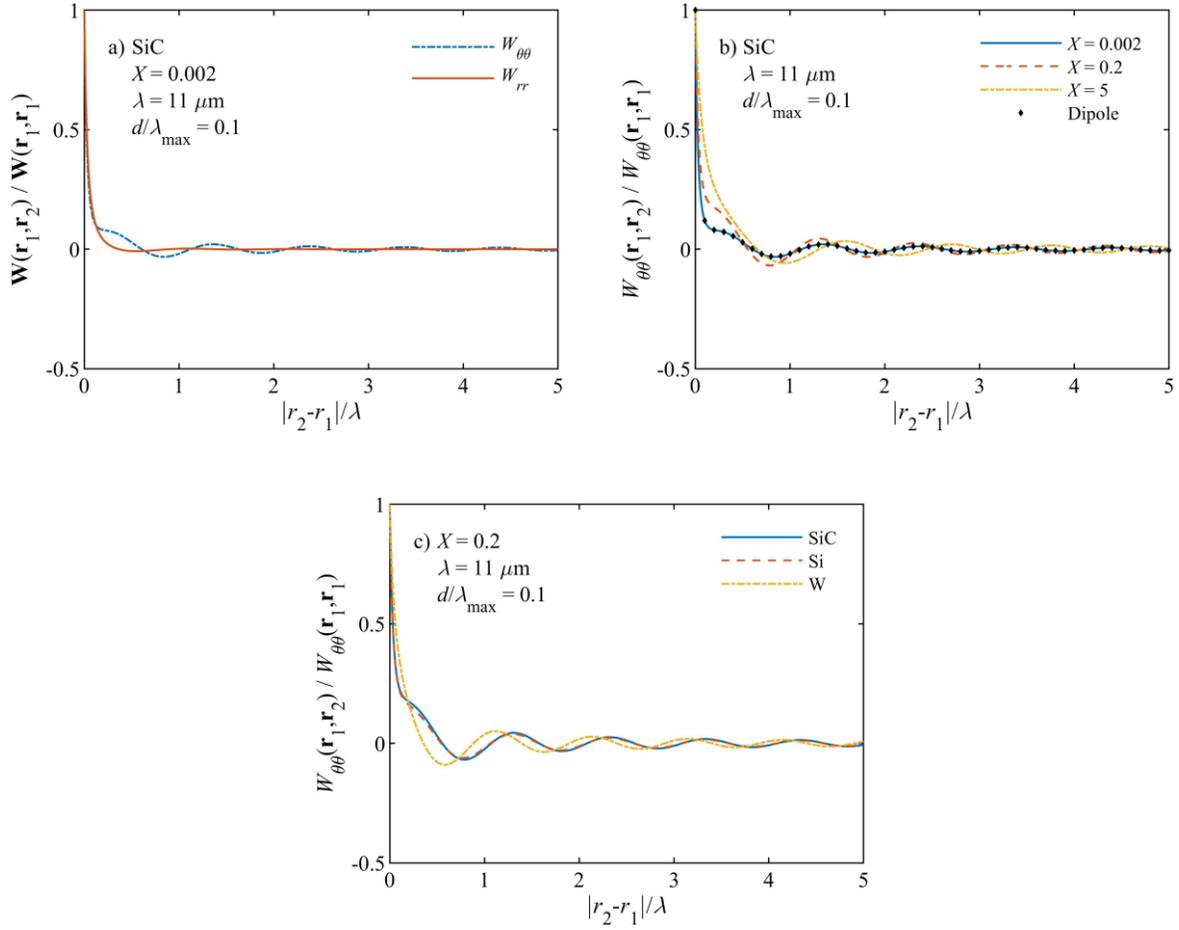

Figure 5 – Same as Fig. 4 for the intermediate near-field regime ($\frac{d}{\lambda_{max}} = 0.1$).

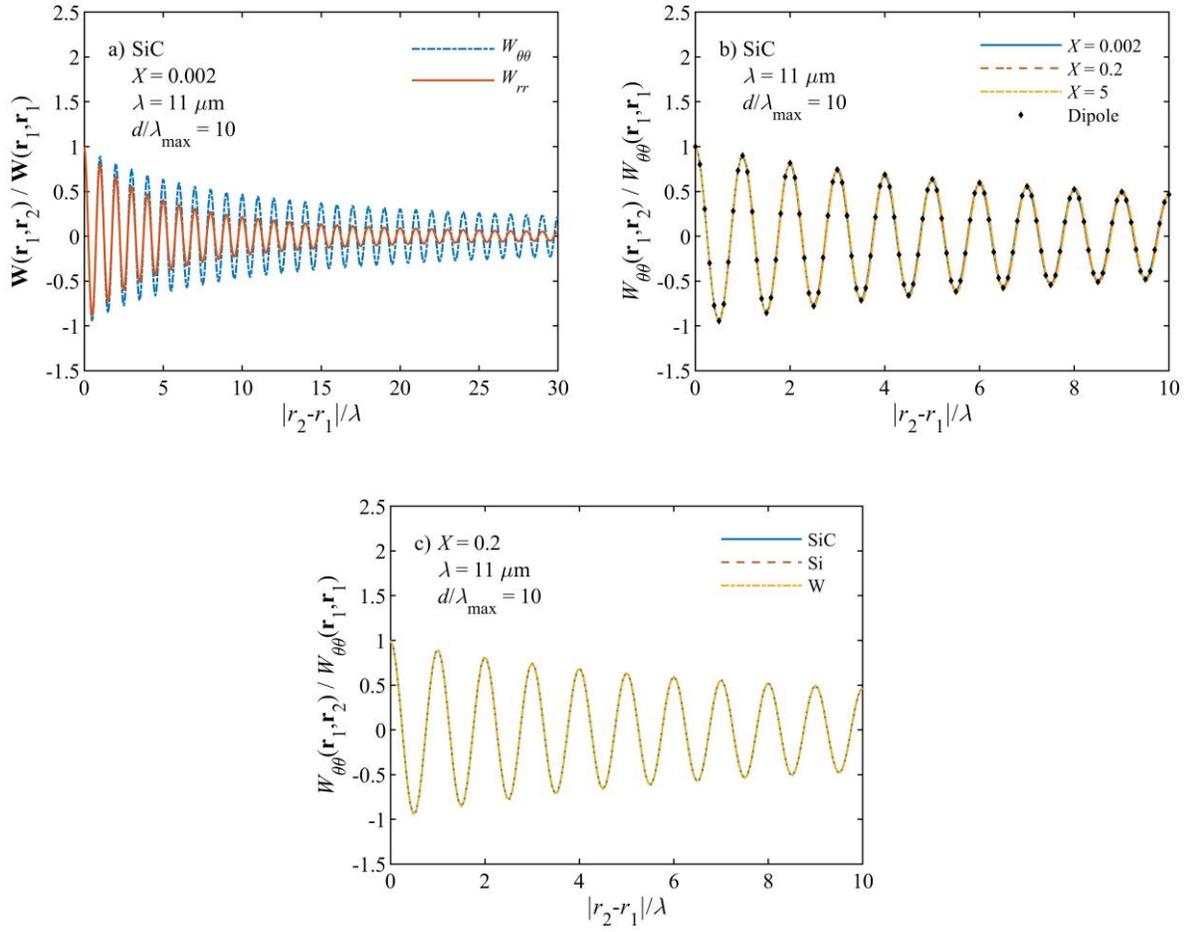

Figure 6 – Same as Fig. 4 for the far-field regime ($\frac{d}{\lambda_{max}} = 10$).

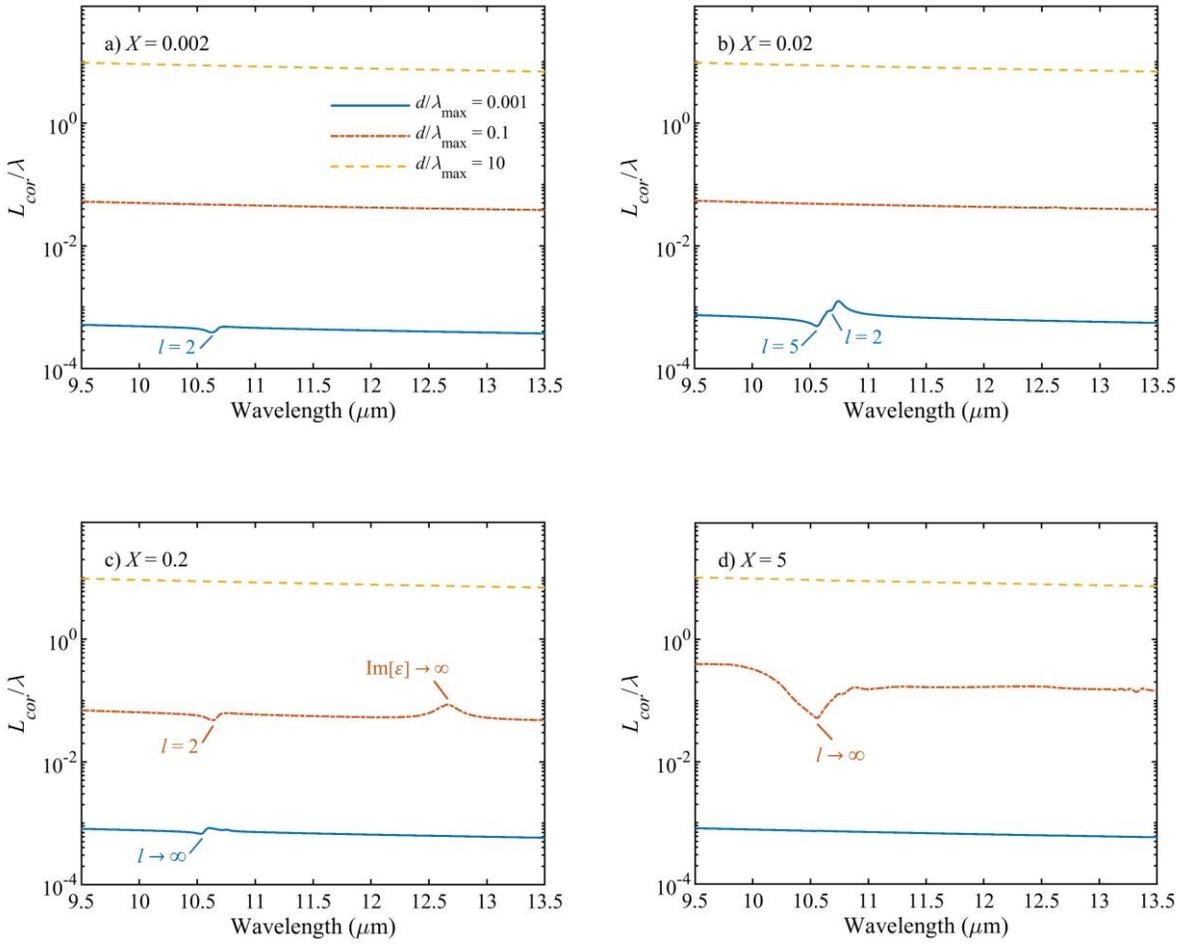

Figure 7 – Normalized correlation length versus wavelength for a SiC sphere with (a) $X = 0.002$, (b) $X = 0.02$, (c) $X = 0.2$, and (d) $X = 5$.

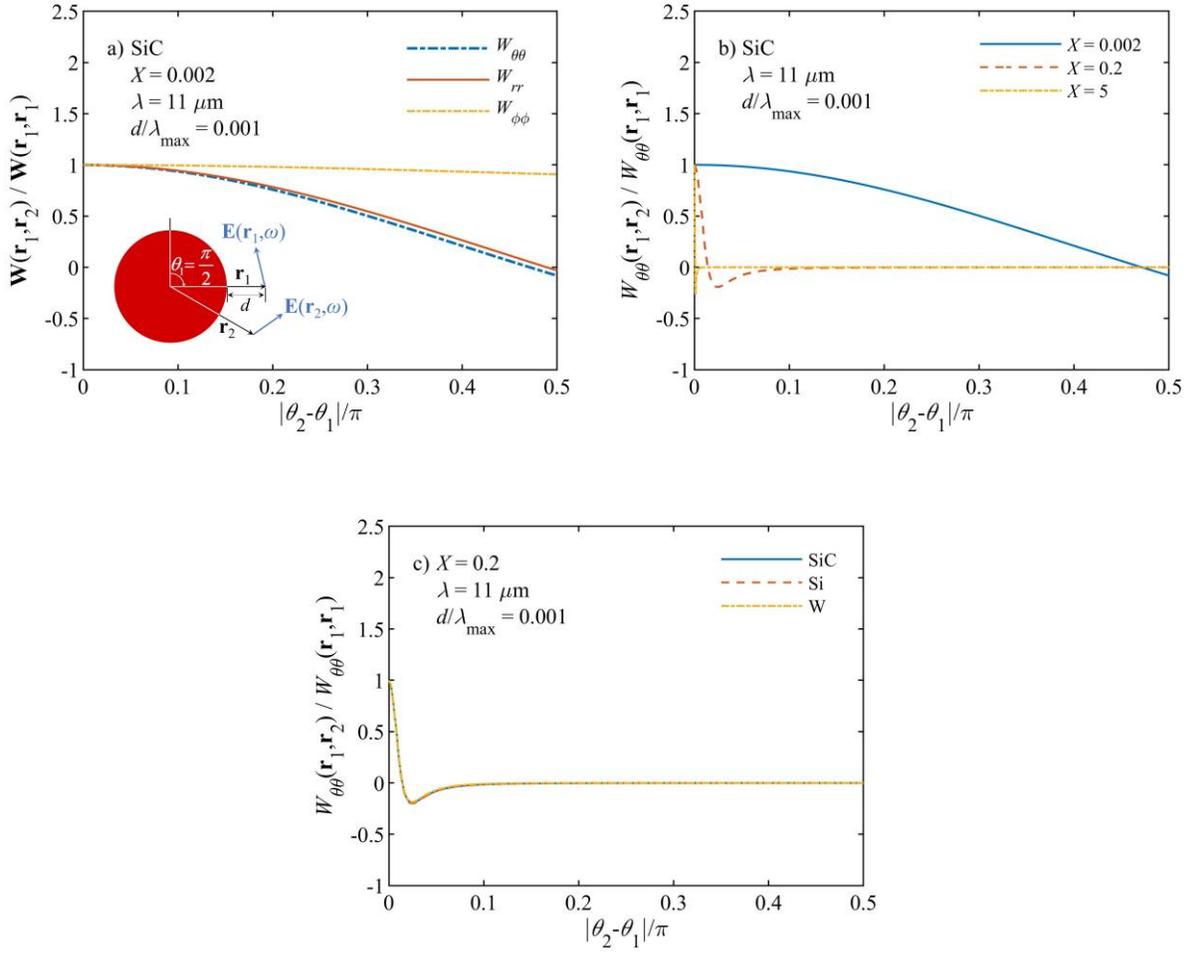

Figure 8 – Normalized correlation function along polar direction in the extreme near-field regime ($\frac{d}{\lambda_{max}} = 0.001$) versus $\frac{|\theta_2 - \theta_1|}{\pi}$ at $\lambda = 11$ µm. (a) Normalized $W_{\theta\theta}$, $W_{rr}$ and $W_{\varphi\varphi}$ for a SiC sphere with $X = 0.002$. (b) Normalized $W_{\theta\theta}$ for SiC spheres with various size parameters. (c) Normalized $W_{\theta\theta}$ for SiC, Si, and W spheres with $X = 0.2$.

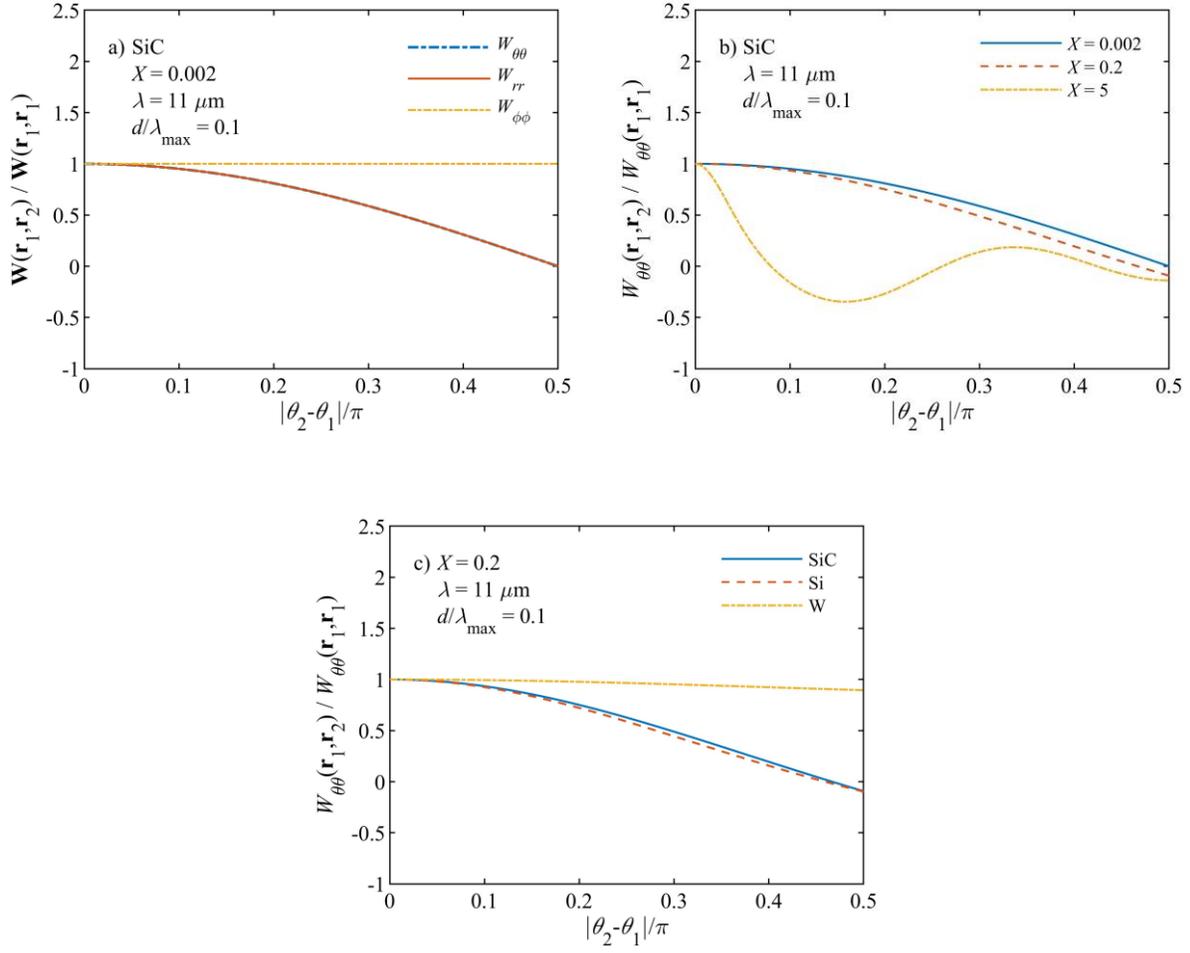

Figure 9 – Same as Fig. 8 for the intermediate near-field regime ($\frac{d}{\lambda_{max}} = 0.1$).

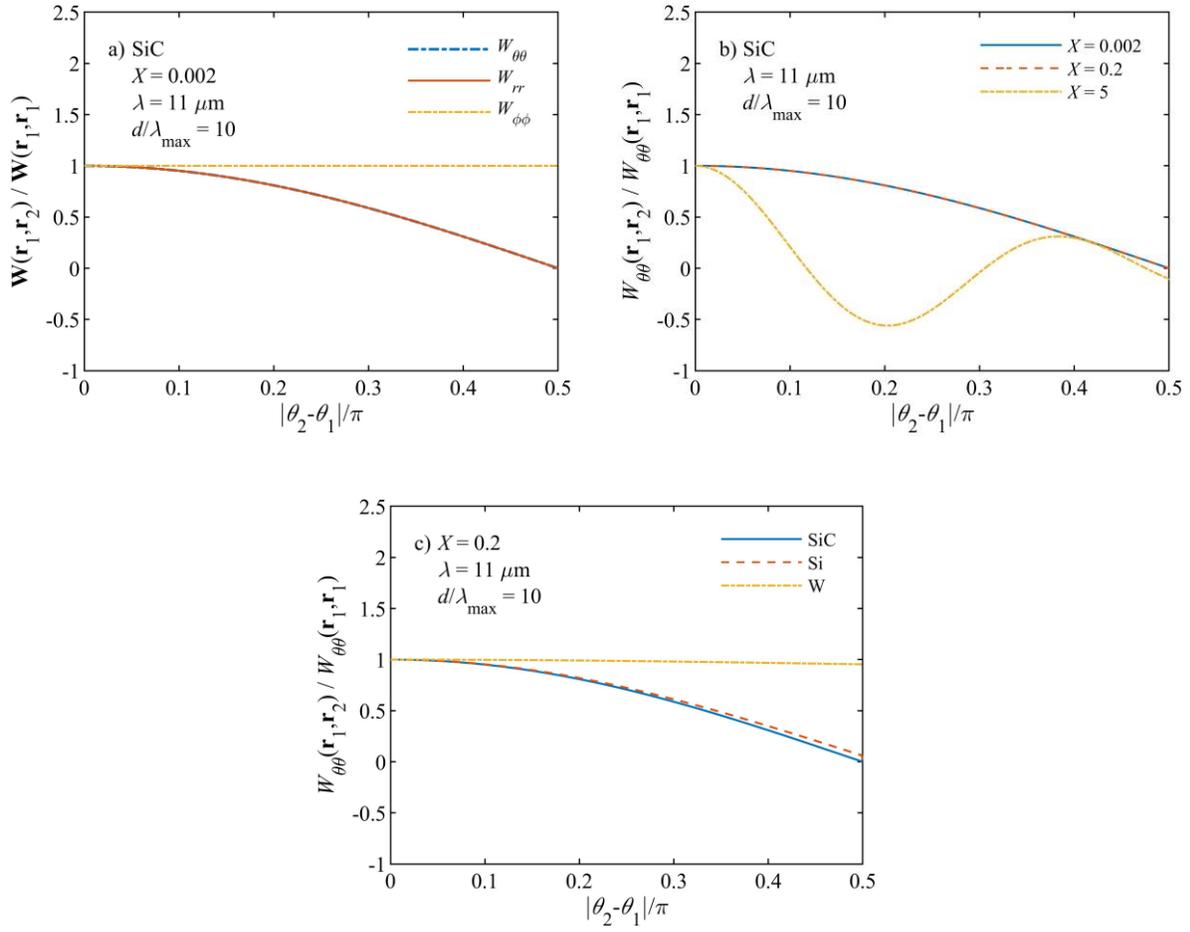

Figure 10 – Same as Fig. 8 for the far-field regime ($\frac{d}{\lambda_{max}} = 10$).

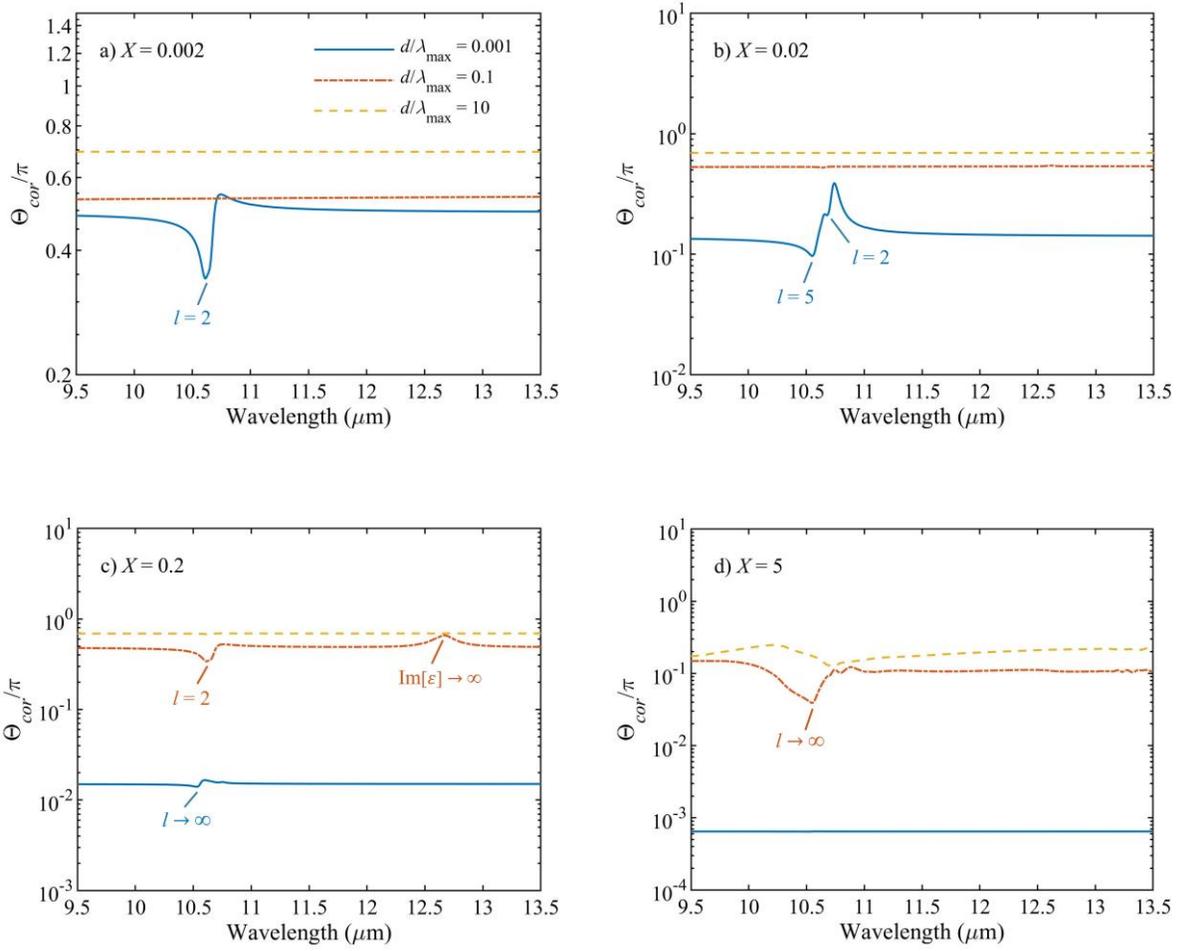

Figure 11 – Normalized correlation angle versus wavelength for a SiC sphere with (a) $X = 0.002$, (b) $X = 0.02$, (c) $X = 0.2$, and (d) $X = 5$.